\documentclass[fleqn,usenatbib]{mnras}

\usepackage{graphicx}	
\usepackage{amsmath}	
\usepackage{amssymb}	

\title[Cosmological DCBH formation sites hostile for their growth]
{Cosmological DCBH formation sites hostile for their growth}

\author[S. Chon et al.]{
Sunmyon Chon $^{1}$\thanks{E-mail: sunmyon.chon@astr.tohoku.ac.jp},
Takashi Hosokawa $^{2}$,
and Kazuyuki Omukai $^{1}$
\\
$^{1}$Astronomical Institute, Graduate School of Science, Tohoku University, Aoba, Sendai 980-8578, Japan\\
$^{2}$Department of Physics, Graduate School of Science, Kyoto University, Sakyo, Kyoto 606-8502, Japan
}

\date{Accepted XXX. Received YYY; in original form ZZZ}

\pubyear{2020}

\begin{document}

\label{firstpage}
\pagerange{\pageref{firstpage}--\pageref{lastpage}}
\maketitle

\begin{abstract}
The direct collapse (DC) is a promising mechanism that provides massive seed black holes (BHs) with $\sim 10^{5}~M_{\odot}$ in the early universe. To study a long-term accretion growth of a DCBH thus formed, we perform cosmological radiation-hydrodynamics simulations, extending our previous work where we investigated its formation stage. With a high spatial resolution down below the Bondi radius, we show that the accretion rate onto the BH is far below the Eddington value. Such slow mass growth is partly because of the strong radiative feedback from the accreting BH. Moreover, we find that the BH has a large velocity of $\sim 100~{\rm km~s^{-1}}$ relative to the gas after it falls into the first galaxy, which substantially reduces the accretion rate. The latter effect stems from the fact that the DCBHs form in metal-free environments typically at $\sim 1~$kpc from the galaxy.
The BH accelerates as it approaches the galactic center due to the gravity. The relative velocity never damps after that, and the BH does not settle down to the galactic center but continues to wander around it. An analytic estimate predicts that 
the DCBH formation within $\sim 100$~pc around the galactic center is necessary to decelerate the BH with dynamical friction before $z=7$. Since metal enrichment with $Z \sim 10^{-5} - 10^{-3}~Z_\odot$ is expected in such a case, the formation of DCBHs in the metal-poor environments is preferable for the subsequent rapid growth.
\end{abstract}

\begin{keywords}
(galaxies:) quasars: supermassive black holes -- stars: Population III
\end{keywords}

\section{Introduction}
Recent observations have revealed a large population of the Supermassive Black Holes (SMBHs) at the early universe with $z \gtrsim 6$
\citep[e.g.][]{Mortlock+2011, Wu+2015, Matsuoka+2016, Onoue+2019}.
The most distant BH found so far is located at $z = 7.54$ \citep{Banados+2018}, which corresponds to the cosmic age of $0.7~$Gyr.
For a seed BH to grow in mass via the Eddington-limited accretion, its initial mass needs to be $\gtrsim 10^{3}~M_{\odot}$. This mass scale is much greater than those of the BHs in the local universe observed via X-rays \citep[e.g.][]{Casares+2014} and recently by gravitational waves \citep[e.g.][]{GW150914}, which ranges from several to several tens solar masses.


The direct collapse (DC) model provides one possible pathway to form SMBHs, in which BHs with $\gtrsim 10^{5}~M_{\odot}$ are seeded in the primordial environment \citep[e.g.][]{BL2003, Volonteri2010, Inayoshi+2019}.
The DC supposes that the collapse of a primordial cloud is delayed until the host halo becomes sufficiently massive as $10^{7}$ -- $10^{8}~M_{\odot}$ by some physical processes, such as the irradiation by strong UV radiation \citep[e.g.][]{Omukai2001, Shang+2010, Latif+2013, Sakurai+2015, Chon+2016, Dunn+2018, Luo+2018, Matsukoba+2019, Suazo+2019}, compression by strong shocks \citep{Inayoshi+2012, Inayoshi+2014}, 
or exposure to the supersonic streaming motion of baryons relative to the dark matter \citep{Hirano+2017b}.
If such extreme conditions are met, the massive primordial cloud collapses, potentially leading to the formation of supermassive stars (SMSs) with $\sim 10^{5}$--$10^{6}~M_{\odot}$. 
Such SMSs finally collapse into BHs with similar masses as their progenitors',
once the general relativistic instability operates 
\citep[e.g.][]{Iben63, Chandra64, Shibata+2002, Umeda+2016, Uchida+2017, Woods+2017, Woods+2020}.


The efficient mass accretion is a possible process to grow the seed BHs to the observed SMBHs. However, whether that occurs or not is still under debate. For example, cosmological simulations with large box sizes conclude that SMBHs exceeding $10^9~M_\odot$ form beyond $z = 6$ as efficient mass accretion at about the Eddington rate continues for a billion years \citep[e.g.][]{Li+2007, DiMatteo+2012, Feng+2014, Smidt+2018}. These authors, however, did not resolve the spatial scale smaller than the Bondi radius, and just adopted the prescription that the efficient Eddington accretion continues.  Since the Bondi rate is proportional to $\rho / c_\text{s}^3$, where $\rho$ is the gas density and $c_\text{s}$ is the sound speed, the actual accretion rate onto the BH should depend on the thermal state of the unresolved gas. The efficient accretion is only possible when the dense and cold gas resides in the unresolved Bondi region, which must be verified by further studies \citep[e.g.][]{Beckmann+2019}.


Indeed, simulations resolving the vicinity of the BH have questioned the efficient mass accretion onto the BHs. For instance, intense star formation near the BH injects a large amount of mechanical energy into the surrounding gas, which easily destroys the cold and dense clouds \citep[e.g.][]{Dubois+2015}. Furthermore, the intense radiation from the accreting BHs also heats the surrounding gas up. These effects reduce the accretion rates far below the Eddington value, both for the BHs with $M_\text{BH}=100~M_{\odot}$ \citep[e.g.][]{Johnson+2007,Alvarez+2009,Milos+2009,PR2011,Jeon+2012} and with $M_\text{BH}=10^{5}~M_{\odot}$ \citep[e.g.][]{Johnson+2011,Aykutalp+2013,Aykutalp+2014}. Recent cosmological simulation by \citet{Latif+2018}, which well resolves the Bondi radius ($\sim$pc), has shown that the seed BH with $10^5~M_\odot$ grows by accretion only by a few \% for 320 million years after $z=12$, while the Pop III and II star formation continues in the same halo uninterruptedly.


A limitation of the previous studies comes from a lack of knowledge on the DCBH forming environments. They did not follow the formation of a DCBH but put it arbitrarily by hand at the center of a massive halo \citep[e.g.][]{DiMatteo+2012, Latif+2018}. However, the DCBHs do not form at the centers of massive star-forming galaxies. The DC model predicts that the BHs should appear at $10^{2}$ -- $10^{3}~$pc away from the center of such galaxies, to avoid the metal enrichment caused by the active star formation \citep{Chon+2016, Regan+2017, Wise+2019}.
In this case, there should be some time-lag until the BH eventually settles down to the galactic center. Although observations show that at $z \lesssim 1$ SMBHs are located very close to the galactic centers 
\citep[e.g.][]{Shen+2019}, this is not necessarily applicable to their counterparts in the early universe. It is still uncertain when the seed BH starts to grow to an SMBH residing at the galactic center. To know it, we need to follow the BH growth starting from their birth.


In this paper, we investigate how efficiently a DCBH accretes the gas considering its formation site with cosmological simulations.
We have previously identified DCBH-forming clouds in a cosmological volume in \citet{Chon+2016}.
We have also confirmed that the formation of the SMSs takes place in these clouds, through the protostellar accretion stage \citep{Chon+2018}. Extending these previous studies, we further follow the long-term accretion growth of the DCBH, performing cosmological radiation-hydrodynamics simulations. We particularly examine how the specific environment surrounding the DCBH affects its growth.
We show that its birthplace predestines the accretion history onto the BH.
Since the BH initially emerges far from the massive galaxy, it has a high velocity of $\sim 100~{\rm km~s^{-1}}$ relative to the galactic center. The BH does not settle down to the center directly. Instead, it continues to wander around the galactic outskirts where the density is much smaller than at the center.
The accretion rate onto the BH, which is proportional to the Bondi rate, drops in such rarefied environments. This effect further delays the BH accretion growth in addition to the relatively well-studied feedback effects from the nearby stars and BH itself. 


This paper is organized as follows. Section~\ref{sec::methodology} describes the numerical procedure and implementation of the physical process.
We present our main results in Section~\ref{sec::results}. 
We discuss our findings in Section~\ref{sec::discussion} and
summarize our results in Section~\ref{sec::summary}.
Throughout the paper, we adopt the cosmological parameters obtained by \citet{PlanckXVI2014}, where $\Omega_\text{matter} = 0.304$, $\Omega_\text{baryon} = 0.0483$, $\Omega_{\Lambda} = 0.692$, $h = 0.677$, and $\sigma_{8} = 0.8288$.

\section{Methodology} 
\label{sec::methodology}

In this paper, we follow the mass accretion history of a seed BH whose formation process has been studied in a full cosmological context \citep{Chon+2016, Chon+2017, Chon+2018}. We term this BH as a DCBH hereafter as it forms satisfying the standard conditions for the DC model (see Section~\ref{sec::dcbhs}).
We perform an N-body + Smoothed Particle Hydrodynamics (SPH) simulation using {\tt Gadget-2} \citep{Springel2005}. The initial condition is generated at $z = 99$ by {\tt MUSIC} \citep{HahnAble2013}, which employs the second-order Lagrangian perturbation theory. The cosmological initial condition is the same as that used in \citet{Chon+2016}.
We first perform the base simulation with a cosmological volume of $20~h^{-1}~$comoving Mpc on a side employing $128^{3}$ dark matter particles.  We next zoom in a specified region with $1.2~h^{-1}~$comoving Mpc on a side and re-simulate the evolution with much higher resolution with masses of the gas and dark matter (DM) particles of $1.8 \times 10^2$ and $1.0 \times 10^3~h^{-1}M_{\odot}$, respectively. To improve the spatial resolution further, we split the gas particles at $z=27.5$ inside a central region with $60~h^{-1}~$comoving kpc containing a massive halo. After the splitting, the mass of each gas particle is $14~h^{-1} M_\odot$ at the halo center.


Note that in \citet{Chon+2016} and \citet{Chon+2017}, we have modeled the star formation and the associated metal enrichment and feedback processes in a semi-analytical manner. In contrast, we directly solve these processes in the current radiation-hydrodynamics simulation, because they determine the thermal state of the gas that the BH accretes. 
Here, we briefly overview the newly implemented physics in our simulation.

\subsection{Chemistry}

We solve non-equilibrium chemical reaction network for 9 species (e$^{-}$, H, H$^{+}$, He, He$^{+}$, He$^{2+}$, H$_{2}$, H$_{2}^{+}$, and H$^{-}$) with an implicit scheme.
The basic chemical network and reaction rates are taken from \citet{Yoshida+2003, Yoshida+2006}. We further consider the photo-reactions induced by the radiation from stars and accreting BHs (see Section~\ref{sec::radiationfeedback}).


We incorporate radiative cooling processes relevant to the primordial chemical reactions in the gas energy equation.
In addition, we consider the cooling by the fine structure emission lines of [C~{\sc ii}]~158$\mu$m and [O~{\sc i}]~63$\mu$m, which are the dominant processes in the low-metallicity gas cloud at $n \lesssim 10^{3}~\mathrm{cm^{-3}}$ \citep[e.g.][]{Omukai2000, Omukai+2008, Chiaki+2016}.
We assume that all the gas phase C and O are in the form of C~{\sc ii} and O~{\sc i}, respectively, without
solving non-equilibrium chemical network including C and O.
We also assume that their abundance ratios follow those of the solar neighborhood and the absolute values linearly scale with the metallicity $Z$,
that is assigned to each gas particle (see Section~\ref{sec::SNfeedback}).
The radiative cooling rates associated with these fine-structure lines are calculated assuming optically-thin, two level system
\citep{Wolfire+1995, Maio+2007}.

\subsection{Star Formation and Stellar Feedback}
Once the gas density exceeds the threshold value 
$n_\text{SF} = 10^3~\mathrm{cm^{-3}}$,
we insert a star particle representing a newborn star cluster.
We model the Pop~III and Pop~II star formation separately,
depending on the metallicity of the star forming gas.
We also implement the stellar feedback processes 
including the ionizing radiation and supernova (SN) feedback.
Here, we describe our modeling for them. 

\subsubsection{Pop~III star formation}
We insert a Pop III star particle with the fixed mass $100~M_\odot$ \citep[e.g.][]{Hirano+2014}, once a gas particle satisfies the following conditions, (i) the density $n > n_\text{SF}$, and (ii) the metallicity $Z < Z_\text{crit, popII} = 10^{-4}~Z_{\odot}$.
We assume that all the Pop III particles have the same radiative properties as a massive Pop III star with $100~M_{\odot}$ \citep{Schaerer2002}; they radiate at a constant luminosity of $1.2\times10^{40}~\mathrm{erg~s^{-1}}$ with the black-body spectrum with $T_\text{eff} = 10^{5}~$K.
We assume they experience core-collapse SNe with the explosion energy of $10^{51}~$erg after their lifetime of two million years.


Note that we here regard the Pop~III star particle as a Pop~III star cluster, and
we assume that one of the member stars with a few $\times 10~M_\odot$ experiences core-collapse SN 
with given explosion energy. This is consistent with recent numerical results suggesting that a primordial cloud usually hosts multiple high-mass Pop~III stars
albeit their mass distribution still being disputed  \citep[e.g.][]{Susa+2014, Stacy+2016, Susa2019, Chon&Hosokawa2019, Sugimura+2020}.

\subsubsection{Pop~II star formation}
Pop~II stars are formed from a cold and metal-enriched gas. In practice, we insert Pop~II star particle when 
a gas particle satisfies the following criteria,
(i) the density $n > n_\text{SF} = 10^3~\mathrm{cm^{-3}}$,
(ii) the metallicity $Z > Z_\text{crit, popII} = 10^{-4}~Z_{\odot}$, and
(iii) the temperature $T < T_\text{crit} = 1000~\mathrm{K}$.
We define a star formation time-scale $t_\text{SF}$ as
\begin{align}
t_\text{SF} = \frac{t_\text{ff}}{\alpha_{*}},
\end{align}
where $t_\text{ff} \equiv (G\rho)^{-1/2}$ is a free-fall time, 
$G$ is the gravitational constant, 
and $\alpha_{*}$ is a star formation efficiency,
where we set $\alpha_{*}=0.07$ \citep{Wise&Cen2009}.
We convert the gas into star particles over the time-scale $t_\text{SF}$ 
according to the following procedure \citep{Okamoto+2008}:
we evaluate the probability that a gas particle is converted into a star particle 
during a given time-step $\Delta t$ as
\begin{align}
p_{*} = \frac{M_\text{part}}{M_{*}} \left [ 1 - \exp \left(-\frac{\Delta t}{t_\text{SF}} \right ) \right ],
\end{align}
where $M_\text{part}$ and $M_{*}$ are the masses of the gas particle and the star particle, respectively.
This star particle represents a star cluster whose mass spectrum follows the Salpeter IMF with the mass range from $0.1$--$100~M_\odot$.
The stellar luminosity and the rate of core-collapse SNe
are calculated by population synthesis code, {\tt STARBURST99} \citep{Leitherer+1999}.

\subsubsection{Radiation Feedback} \label{sec::radiationfeedback}
We include the Lyman-Werner (LW) and ionizing radiation feedback from stars.
The three photo-reactions are considered here, H$_{2}$ photo-dissociation, photo-detachment of H$^{-}$, and photo-ionization of H,
\begin{align}
& \mathrm{H}_{2} + \gamma \longrightarrow 2\mathrm{H}, \label{ReacH2diss} \\
& \mathrm{H}^{-}  + \gamma \longrightarrow \mathrm{H} + \mathrm{e}^{-}, \label{ReacH-diss} \\
& \mathrm{H}       + \gamma \longrightarrow \mathrm{H}^{+} + \mathrm{e}^{-}. \label{ReacHioniz}
\end{align}
Reactions~\ref{ReacH2diss} and \ref{ReacH-diss} reduce the H$_{2}$ abundance,
thereby suppressing the primordial star formation.

We divide the radiation spectrum into two components \citep{Chon+2017}: 
(i) H$_{2}$ dissociating radiation ($h\nu < 13.6~$eV) and (ii) ionizing radiation ($h\nu > 13.6~$eV).
\begin{enumerate}
\item Dissociating radiation \\
We assume the radiation is optically thin since we only focus on low density regime 
($n \lesssim 10^3~\mathrm{cm^{-3}}$).
By summing the contribution from the local radiation sources, 
we can evaluate the spatial distribution of the intensity of the radiation in the LW band as \citep{Agarwal+2012, Johnson+2013},
\begin{align}
J_{21} = \sum_{i} \frac{1}{\pi} \frac{L_{\text{LW}, i}}{\Delta \nu_\text{LW}} \frac{1}{4 \pi r_{i}^{2}},
\end{align}
where $\Delta \nu_\text{LW}$ is the width of the LW band,
$L_{\text{LW}, i}$  is the luminosity of the LW radiation of the point source $i$,
and $r_{i}$ is the distance from the point source at a given point.\\

\item Ionizing radiation \\
We consider photo-ionization of H and the associated photo-heating by the radiation with $h\nu > 13.6~$eV.
To calculate the transfer of the ionizing radiation, 
we evaluate the optical depth $\tau$ for ionizing photons
following a ray-tracing scheme, RSPH \citep[e.g.][]{Kessel-Deynet+2000, Susa2006, Chon+2018}
and attenuate the intensity proportionally to $\exp \left ( - \tau \right )$.
The radiation spectra are assumed to be the black-body with $T_\text{eff} = 10^{5}~$K and $2 \times 10^{4}~$K
for Pop~III and Pop~II sources, respectively.
Here, we do not consider the attenuation and 
re-emission by the dust,
since the mean metallicty in the inter-galactic medium 
is $Z/Z_{\odot} \lesssim 10^{-3}$
and the dust opacity inside the ionized region is smaller than unity 
\citep{Raga+2015}.
\end{enumerate}

\subsubsection{Supernovae and metal enrichment} \label{sec::SNfeedback}
After ending their lives, stars with a given mass range experience core-collapse SNe,
thrust the surrounding gas and scatter heavy element around.
To emulate this, we inject thermal energy into the gas particles 
within $0.5~$comoving kpc from the star particle, 
assuming the injected explosion energy quickly thermalizes \citep{Springel&Hernquist2003, Dubois&Teyssier2008}.
Since our simulation does not spatially resolve 
small hot region with $T > 10^{7}$--$10^{8}~$K,
the temperature of the SN ejecta is underestimated and
the ejecta rapidly cools by bremsstrahlung.
This makes the blast wave dissipates before it propagates through the star-forming cloud \citep{DallaVecchia+2012}.
To circumvent this problem, once the gas particle is heated by an SN, we switch off the gas cooling for $10~$million years,
the typical sound crossing time of the star-forming region \citep[e.g.][]{Governato+2007,Dubois+2012}.

Heavy elements are synthesized inside a star and 
scattered by the SN explosion.
We uniformly distribute the mass of the heavy elements $M_\text{metal}$
to the gas particles located within $1~$comoving kpc from the star particle of explosion.
The heavy elements are supposed to spread around by the dissipation 
due to the unresolved turbulence, etc. as well as the advection.
Instead of solving the diffusion equation,
we smooth out the metal abundance over the smoothing length
and the metallicity $Z_{i}$ of the gas particle $i$
\citep{Tornatore+2007, Wiersma+2009}:
\begin{align}
Z_{i} &\equiv \frac{\rho_{\text{metal},i}}{\rho_{i}}, \\
\rho_{i} &= \sum_{j} m_{j} W(r_{i} - r_{j}, h_{i}), \\
\rho_{\text{metal}, i} &= \sum_{j} m_{\text{metal}, j} W(r_{i} - r_{j}, h_{i}),
\end{align}
where $W$ is the SPH kernel function,  
$m_{j}$, $m_{\text{metal}, j}$, $r_{j}$, and $h_{j}$ 
are the total mass,  
the mass of the heavy elements,
the position, and
the smoothing length
of the SPH particle $j$, respectively.
The index $j$ runs over the particles inside the smoothing length.

We adopt the explosion energy $E_\text{SN}=10^{51}~$erg and the ejected metal mass $M_\text{metal} = 10~M_{\odot}$ 
for the Pop~III star \citep{Umeda&Nomoto2002,Chiaki+2018}.
For the Pop~II star, the SNe continuously occur with time
since a star particle consists of stars with a variety of masses following a given IMF.
We evaluate the SN rate, explosion energy, and the metal mass for every $2~$million years 
using {\tt STARBURST99}.
 
\subsection{Formation, growth, and feedback from DCBHs}
\label{sec::dcbhs}

We insert a BH particle once a gas particle satisfies the condition
required by the DC model \citep[e.g.][]{Omukai2001, Shang+2010, Agarwal+2012, Chon+2016},
where
(i) the metallicity $Z  < Z_\text{crit, DC} = 10^{-6}~Z_{\odot}$, 
(ii) the density $n > n_\text{SF} = 10^{3}~\mathrm{cm^{-3}}$,
(iii) the intensity of the radiation in LW band $J_{21} > J_\text{crit} = 100$ 
\footnote{$J_{21}$ is normalized in the unit of $10^{-21}~\mathrm{erg~cm^{-2} s^{-1} Hz^{-1} str^{-1}}$.}.
Since the mass of the BH finally formed in the DC clouds are quite uncertain \citep{Umeda+2016},
we conduct a simulation with the two different initial BH masses, 
$10^{5}$ and $10^{6}~M_{\odot}$.
We assume that BHs accrete the surrounding gas at a Bondi accretion rate,
but the rate is also limited by the Eddington rate as \citep[e.g.][]{Kim+2011},
\begin{align}
\dot{M}_\text{BH}& = \min \{\dot{M}_\text{Bondi}, \dot{M}_\text{Edd} \},
\end{align}
where 
\begin{align}
\dot{M}_\text{Bondi} &= \frac{4 \pi G^{2} M_\text{BH}^{2} \rho}{(c_\text{s}^{2} + v_\text{rel}^{2})^{3/2}},  \label{eq::MBHacc} \\
\dot{M}_\text{Edd}    &= \frac{4 \pi G M_\text{BH} m_\text{p}}{\epsilon \sigma_\text{T} c} \label{eq::MBHEddington},
\end{align}
where $c_\text{s}$ and $\rho$ are the sound speed and the density of the surrounding gas,
$v_\text{rel}$ is the relative velocity between the BH and the gas,
$m_\text{p}$ is the proton mass, 
$\epsilon = 0.1$ is the radiative efficiency, 
$\sigma_\text{T}$ is the Thomson scattering cross section,
and $c$ is the speed of light.

\begin{figure}
	\centering
		\includegraphics[width=8.5cm]{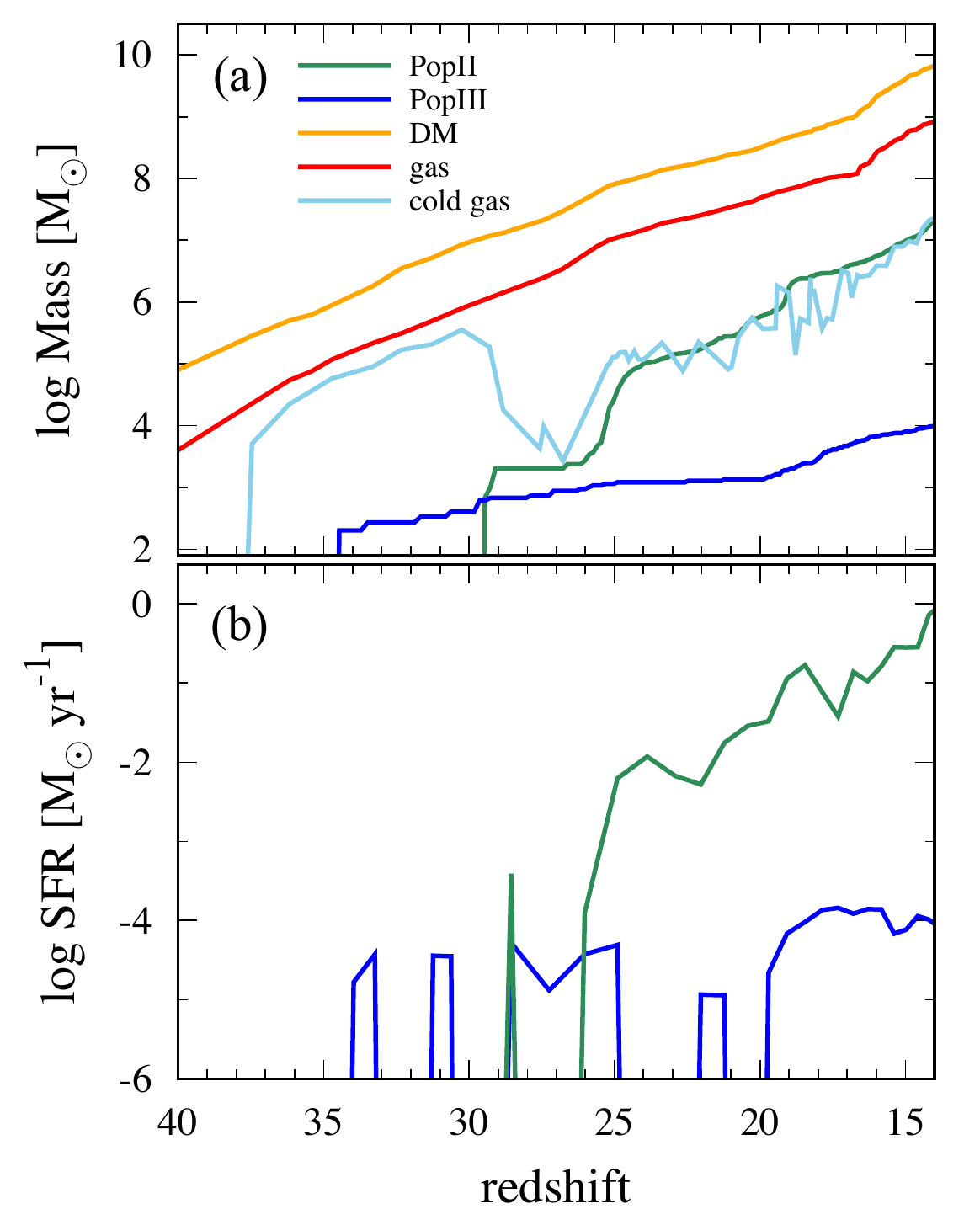}
		\caption{Time evolution of (a) the masses of DM halo (orange), the total gas (red), the cold gas (light blue),
		 Pop~II stars (green), and Pop~III stars (blue), and of
		 (b) SFRs of the Pop~II (green) and Pop~III stars (blue).}
		\label{fig_SF}
\end{figure}

\begin{figure*}
	\centering
		\includegraphics[width=12.5cm]{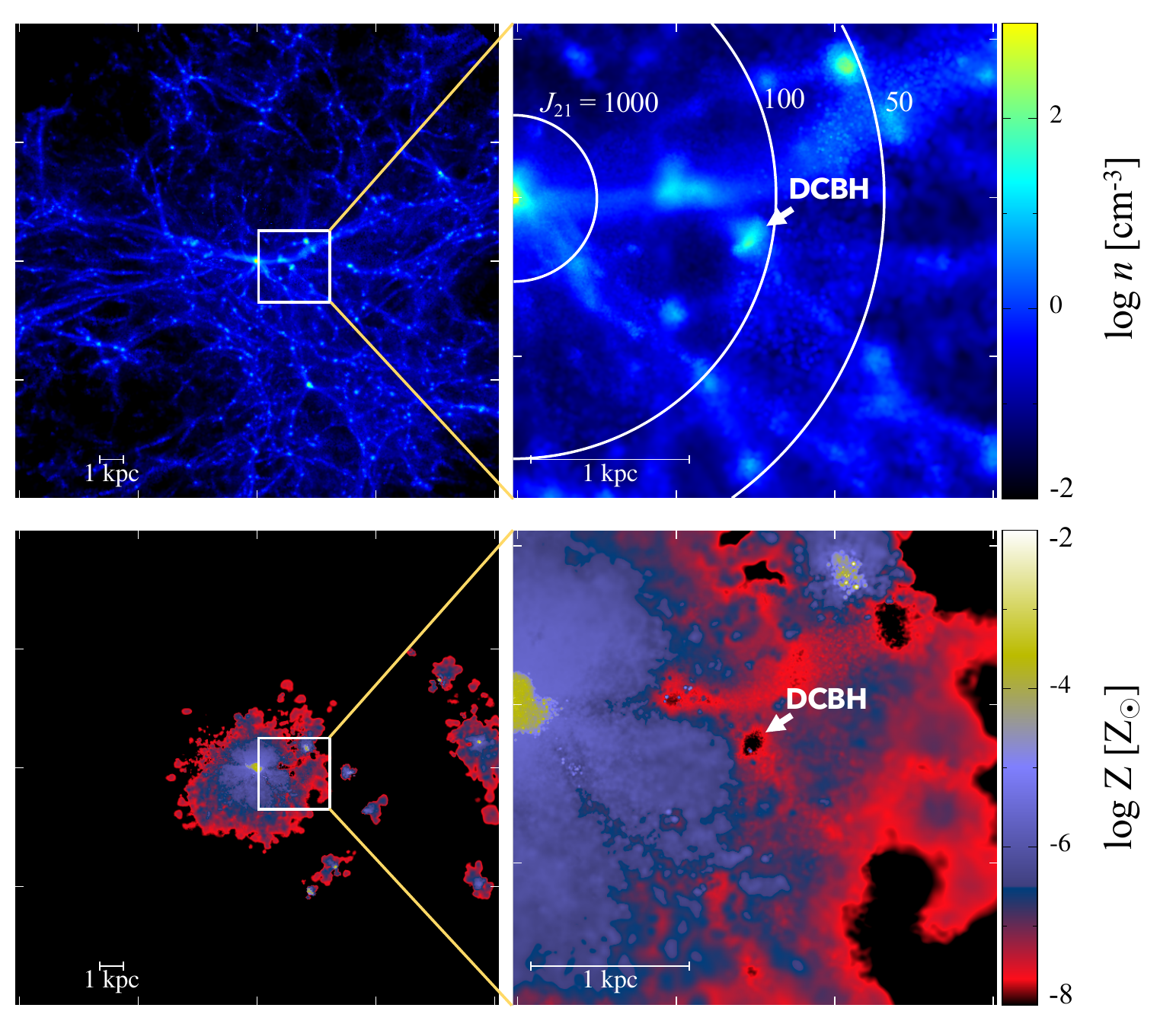}
		\caption{The projected distributions of the density (top) and the gas metallicity (bottom)
		at $z=20.9$, when the BH is forming in our simulated region.
		Panels at right column are the zoom-in views, centered around the cloud 
		which will hatch the seed BH inside.
		The arrow indicates the cloud where the BH will form.
		We overplot the contour of the LW intensity, $J_{21}$, on the right top panel.}
		\label{fig_snapshot_zoom}
\end{figure*}

Note that we do not include the boosting factor to the Bondi accretion rate
which is frequently introduced supposing the presence of unresolved cold and dense gas clumps
in studying BH growth in cosmological simulations
\citep[e.g.][]{Booth&Schaye2009, DiMatteo+2012}.
Our simulation has high enough spatial resolution down to $1$--$10~$pc,
allowing us to resolve the Bondi radius around the BH,
\begin{align}
R_\text{Bondi} &= \frac{2GM_\text{BH}}{{c_\text{s}}^{2}}, \nonumber \\
&= 8.67~\mathrm{pc} \left ( \frac{M_\text{BH}}{10^{5}~M_{\odot}} \right ) \left ( \frac{c_\text{s}}{10~\mathrm{km~s^{-1}}} \right )^{-2}.
\end{align}
Furthermore, since we follow the cooling process of the galactic gas,
we can naturally resolve the contribution from accreting dense and cold clumps.
In our simulation, such cold component has $n \gtrsim 10^3 ~\mathrm{cm^{-3}}$ and $T \lesssim 1000~$K, whose Jeans length is $\lesssim 10~$pc
\citep[e.g.][]{Booth&Schaye2009}.


We also consider luminosity of the accreting BH $L_\text{BH}$, for which a fraction $\epsilon$ of the rest mass energy of the gas is liberated as radiation,
\begin{align}
L_\text{BH} = \epsilon \dot{M}_\text{BH}c^{2}.
\end{align}
We assume non-thermal radiation from the standard BH accretion disk with the spectrum obeying a power law $L_{\nu} \propto \nu^{-1.5}$, where the minimum and maximum energies are $h\nu_\text{min}=13.6~$eV and $h\nu_\text{max}=100~$keV, respectively \citep{Milos+2009b}.
We solve the transfer of the radiation and evaluate the ionization and the associated heating rate, as described in Section~\ref{sec::radiationfeedback}.


\section{Results} 
\label{sec::results}
\subsection{Assembly of first galaxy and DCBH formation}

We first focus on the emergence of the first galaxy, around which the DCBH forms and grows by accreting the gas. An intense radiation field provided by the galaxy is not only necessary for the DCBH formation, but also crucial for the subsequent mass growth as it affects the gas that the BH potentially accretes. We here briefly describe how the first galaxy forms and dominates the surrounding environment.


Fig.~\ref{fig_SF} shows (a) the assembly history of the first galaxy 
and (b) the star formation rates of the Pop~II (green) and Pop~III stars (blue). 
The Pop~III star formation begins at $z \sim 35$ inside a minihalo with a mass of $\sim 10^{6}~M_{\odot}$. After the stellar lifetime of two million years, heavy elements scatter around along with the blast wave powered by the core-collapse SN. The Pop~II star formation starts at $z \sim 29$, when the ejected metals accumulate at the halo center \citep[e.g.][]{Chiaki+2018}. They emit the ionizing radiation and heat the surrounding gas.
The light-blue line in Fig.~\ref{fig_SF}(a) indicates that the amount of the cold gas decreases just after the initial bursty formation of Pop~II stars. 
Pop II star formation occurs continuously after $z \sim 26$ with increasing rate of $0.01$ -- $0.1~M_{\odot}~\mathrm{yr^{-1}}$, indicating the emergence of the first galaxy. During this epoch, the halo mass reaches $10^{8}~M_{\odot}$ and the halo becomes able to gravitationally bind the gas heated by the stellar ionizing radiation. The Pop~II formation rate reaches $\sim1~M_{\odot}~\mathrm{yr^{-1}}$ at $z=14$.

\begin{figure*}
	\centering
		\includegraphics[width=18.cm]{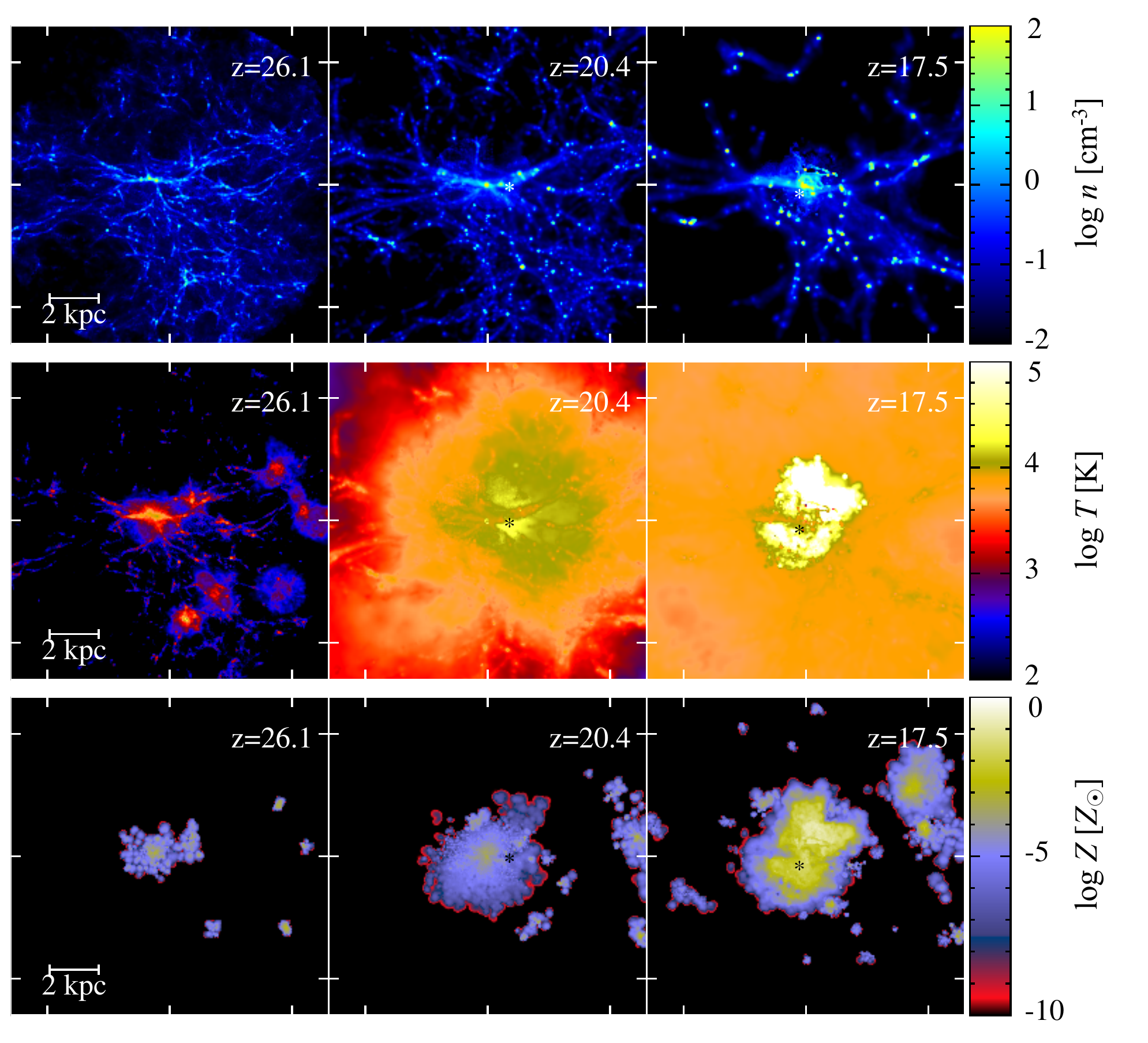}
		\caption{The projected distributions of the density (top), temperature (middle), and metallicity (bottom row)
		at three different epochs of $z=26.1$ (left), $20.4$ (middle), and $17.5$ (right column).
		The asterisks shown in the panels at the middle and right columns indicate the position of the DCBH.}
		\label{fig_snapshot}
\end{figure*}

Fig.~\ref{fig_snapshot_zoom} shows the distributions of the density (top) and the metallicity (bottom) when the DCBH appears at $z=20.9$. The arrow indicates the birthplace of the DCBH, which is located at $\sim 1~$kpc away from the luminous galaxy.
The DCBH birthplace is the same as in our previous studies, where we modeled the metal enrichment and stellar feedback processes in a different semi-analytic way.
The LW intensity at this point is $J_{21} \sim10^{2}$--$10^{3}$, exceeding the critical value required for the DC to take place. The bottom panel shows the metallicity distribution, indicating that the cloud remains almost primordial in composition. The SNe activity in the central galaxy scatters heavy elements around, and the metallicity reaches $\sim 10^{-5}~Z_{\odot}$ on average even at the distance of $\sim$~kpc away from it. Nevertheless, we observe that a dense central part of the cloud avoids enrichment by the heavy elements, being protected by the dense filaments and envelopes, where the metal mixing is inefficient  \citep[e.g.][]{Cen&Riquelme2008, Smith+2015}. We follow the evolution of this DCBH hereafter.

\begin{figure}
	\centering
		\includegraphics[width=8.cm]{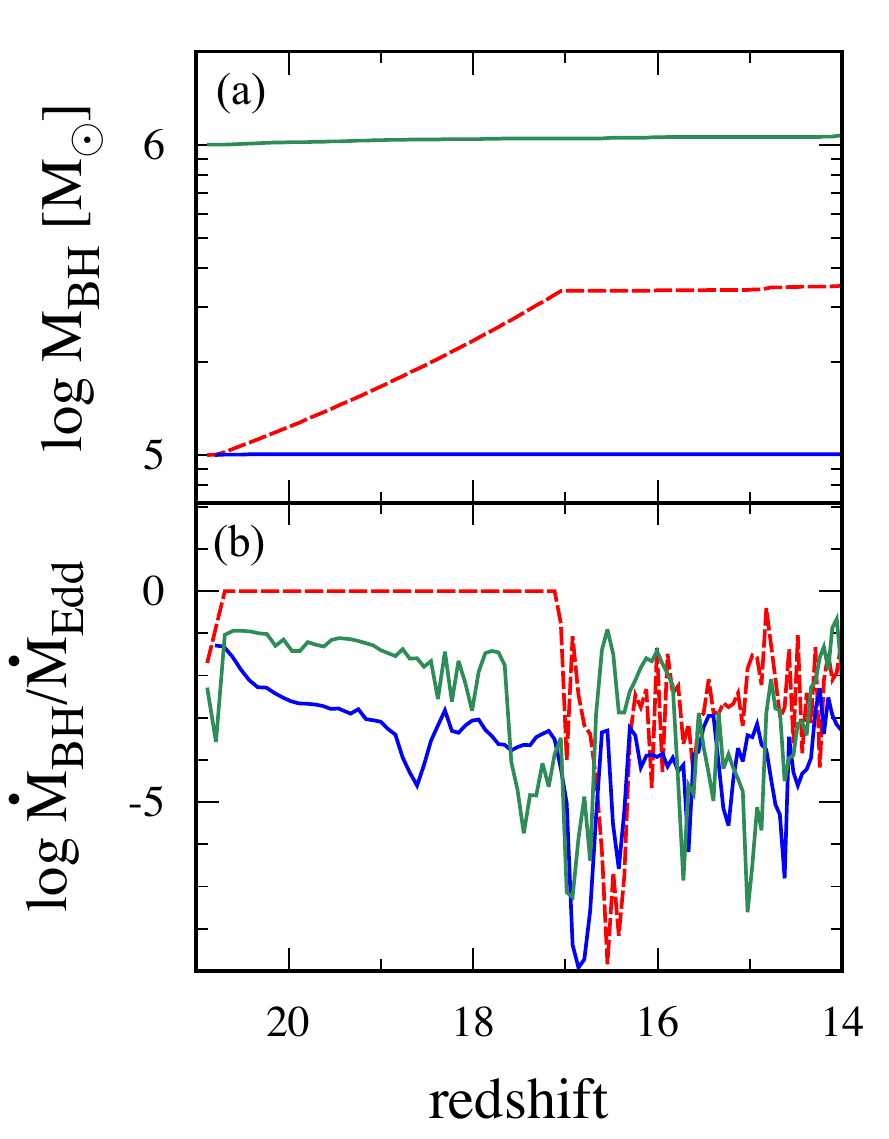}
		\caption{Time evolution of (a) the BH mass and 
		(b) the mass accretion rate onto the BH normalized by the Eddington accretion rate.
		The initial BH mass is assumed to be $10^{5}$ (blue and red) and $10^{6}~M_{\odot}$ (green).
		Dashed line shows the results when we turn off the radiation feedback from the accreting BHs.
		}
		\label{fig_BH_growth}
\end{figure}

\begin{figure}
	\centering
		\includegraphics[width=8.cm]{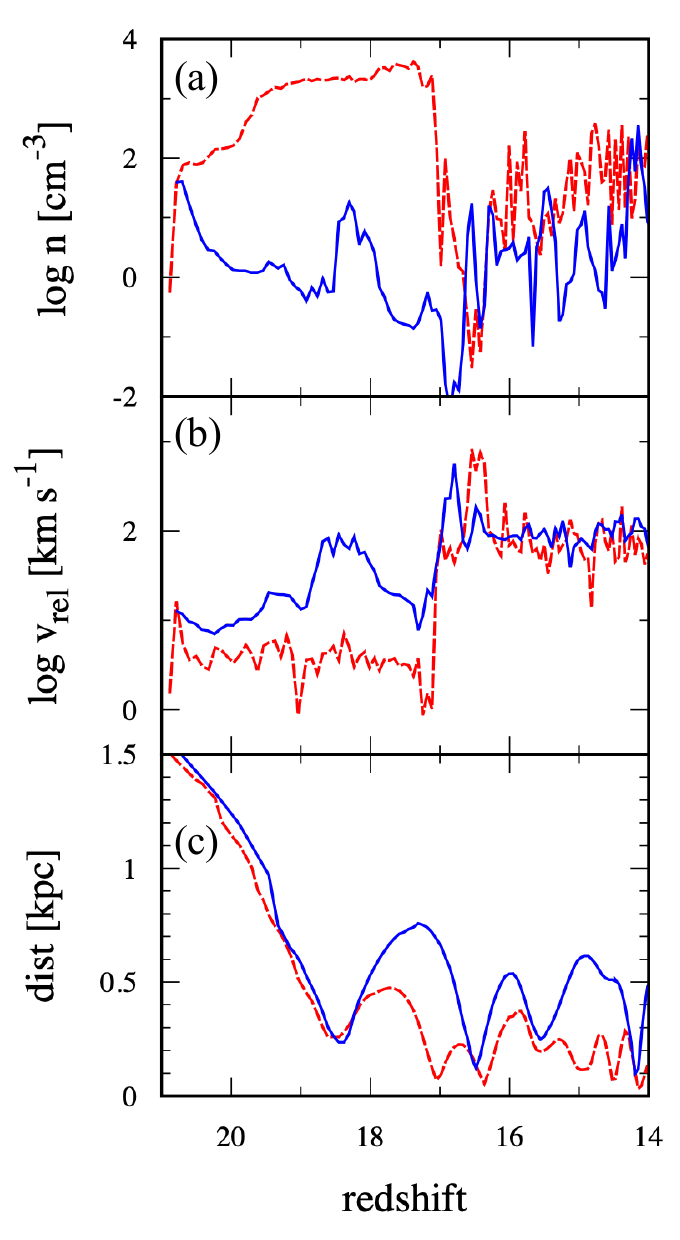}
		\caption{Time evolution of (a) the gas density around the BH,
		(b) the relative velocity between the BH and the surrounding gas, and
		(c) the distance between the BH and the galaxy center.
		Here, we assume the initial BH mass of $10^{5}~M_{\odot}$.
		The solid and dashed lines represent the results with and without
		the radiation feedback from the accreting BH, respectively.
		}
		\label{fig_gas_props_around_BH}
\end{figure}

\begin{figure*}
	\centering
		\includegraphics[width=16.cm]{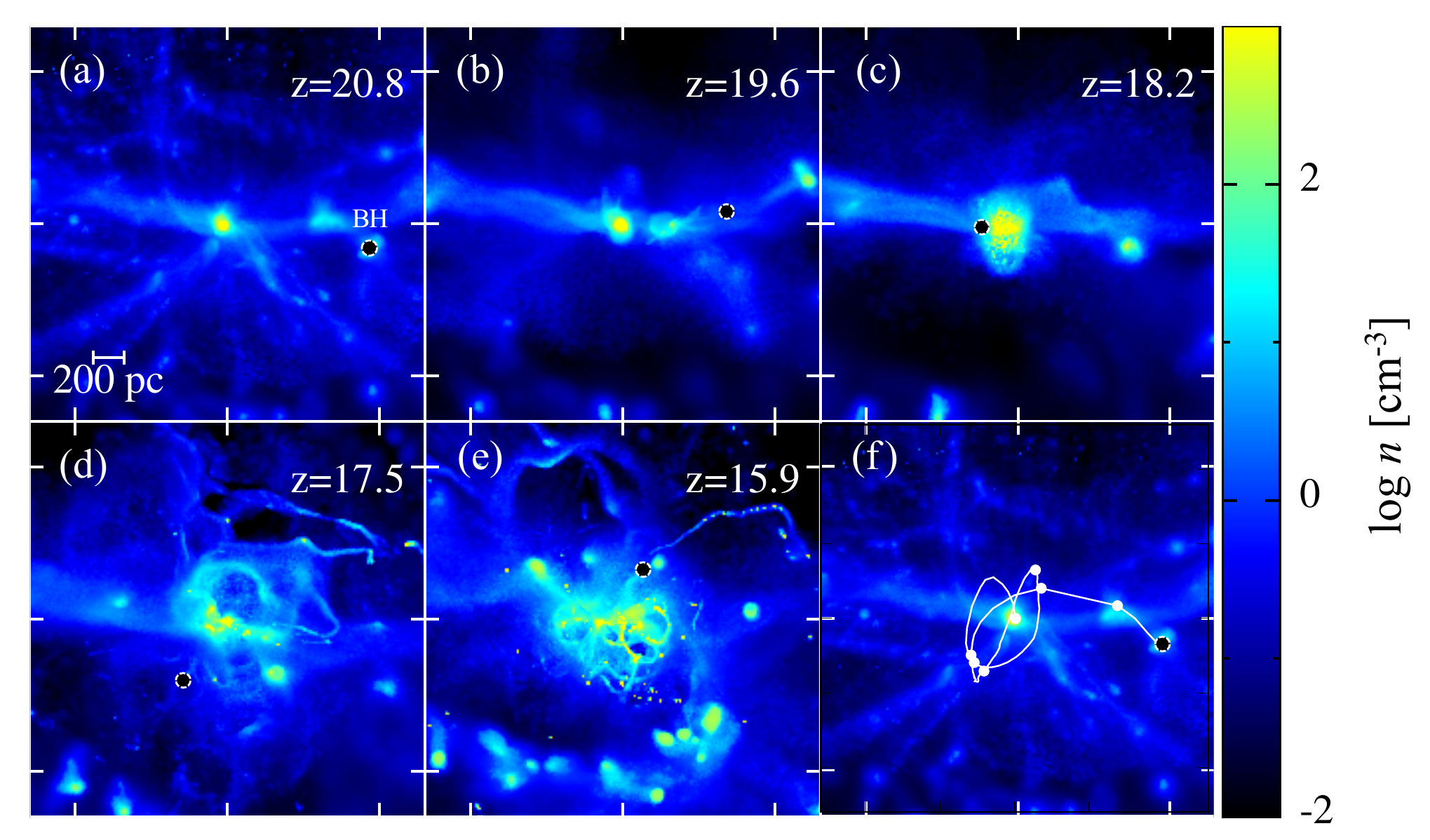}
		\caption{(a)--(e) The projected density distribution around the galaxy center 
		at $z=20.8$, $19.6$, $18.2$, $17.5$, and $15.9$. The circle bounded by the white color shows
		the position of the BH. 
		(f) The trajectory of the BH overplotted on the density distribution at $z=20.8$. 
		The asterisk shows the initial position and the bold circles show the positions of the moving BH 
		at $z=20$, $19$, $18$, $17$, $16$, $15$, and $14$.}
		\label{fig_BH_migration}
\end{figure*}

\subsection{Inefficient mass growth of DCBHs}

We perform simulations with different choices of the initial DCBH masses, 
$10^{5}$ and $10^{6}~M_{\odot}$. To further test the importance of the radiation feedback from the accreting BHs, we also follow the BH growth with no feedback for the model with the initial mass of $10^{5}~M_{\odot}$. We study the BH growth histories for the three different models in total. 


Fig.~\ref{fig_snapshot} shows the projected distributions of the gas density (top), temperature (middle), and metallicity (bottom) for the case where we assume the initial DCBH mass of  $10^{5}~M_{\odot}$ and consider the radiative feedback from the BH.
Panels at the left column correspond to the snapshots at $z=26.1$, when the continuous Pop~II star formation begins. Around the forming first galaxy, the ionized regions start developing and the gas temperature increases to several thousand K. Pop II clusters, locating at the center of the panel, create the largest ionized region at this epoch. The extent of the ionized gas is still small $\lesssim 1~$kpc and most of the clumps in the simulated region
avoid photo-heating and thus photo-evaporation \citep{Chon+2017}. 
Panels at the middle column correspond to the snapshots at $z=20.4$, just after the DCBH is formed. Since the BH growth rate is initially close to the Eddington rate (see Fig.~\ref{fig_BH_growth}b), the intense radiation from the accreting BH fully ionizes the entire simulated region. The asterisk represents the position of the DCBH, around which the temperature exceeds $10^{4}~$K owing to energetic photons emitted from the BH accreting disk. 
Panels at the right column show that the radiative feedback from the accreting BH dramatically changes the density distribution. Filaments and small clumps are photo-evaporated, and only dense clumps survive. We can also see that the intense SNe feedback injects a large amount of energy into the gas inside the galaxy, and the gas temperature suddenly increases up to $10^{5}~$K.


Fig.~\ref{fig_BH_growth} presents (a) the mass evolution of the BH, and (b) the mass accretion rate onto it normalized by the Eddington rate. Solid lines represent the cases with the initial BH mass of $10^{5}$ (red) and $10^{6}~M_{\odot}$ (green) 
while the dashed line shows the results without feedback. 
The solid lines indicate that the radiative feedback from the BH is powerful enough to almost halts the mass growth. The BH radiation heats the surrounding gas and reduces the accretion rate by two to three orders of magnitude.
Until $z=14$, which is the end of our simulations, the BH only grows by $0.6\%$ and $7\%$ in mass for the models with the initial BH masses of $10^{5}$ and $10^{6}~M_{\odot}$, 
respectively.


Fig.~\ref{fig_gas_props_around_BH} shows 
(a) the mean gas density around the BH $n$,
(b) the relative velocity between the BH and the surrounding gas $v_\text{rel}$, and, (c) the separation between the BH and the galaxy center for the models with (solid) and without the radiative feedback from BHs (dashed line) for the initial BH mass of $10^{5}~M_{\odot}$. 
We here evaluate $n$ by averaging the density over the spherical region with a radius of $50~$pc from the BH.
Recall that above two parameters, $n$ and $v_\text{rel}$, are important in determining the accretion rate as the Bondi rate is proportional to $n v_\text{rel}^{-3}$. 
The BH forms within a dense cloud with $n\sim100~\mathrm{cm^{-3}}$,
  and grows efficiently by accretion.
Strong radiation is emitted by liberating the gravitational energy of the massive accreting gas flow, so that the surrounding gas is heated up to the temperature higher than $10^{4}~$K.
This makes the gas escape from the host halo of the BH (so-called photo-evaporation), and the surrounding gas density gradually decreases with time \citep[e.g.][]{Johnson+2011}. The BH growth rate consequently decreases. Moreover, after the gas escape, the relative velocity $v_\text{rel}$ increases to $30$--$100~\mathrm{km~s^{-1}}$, which further reduces the accretion rate. Although the BH encounters with a dense filament with $n\sim10~\mathrm{cm^{-3}}$ at $z=18$--$19$, the accretion rate stays far below the Eddington rate as a result of
high relative velocity between the BH and filament, $100~\mathrm{km~s^{-1}}$.

Even if the radiation feedback is turned off by hand,
the accretion rate suddenly drops by several orders of magnitude at $z \sim 17$
after the initial rapid growth at the Eddington rate (Fig.~\ref{fig_BH_growth}b, dashed line). 
This indicates that the mass growth is prevented not by the radiation from the BH,
but mainly by the SN activity at the galaxy center.
Fig.~\ref{fig_gas_props_around_BH}(b) clearly explains why the accretion rate onto the BH decreases.
The gas density around the BH is initially very high as  $n\sim 10^{3}$--$10^{4}~\mathrm{cm^{-3}}$ and $v_\text{rel} \lesssim c_\text{s}$,
making the accretion rate close to the Eddington.
This efficient growth ceases at $z\sim 17$, when the BH reaches a region with $\sim 100~$pc from the galaxy center.
Around this epoch, the SN blast wave disrupts the high-density gas surrounding the BH and reduces the gas density.
In addition to this, the gas in the central region becomes highly turbulent, as the SNe inject a large amount of energy around (right panel of Fig.~\ref{fig_snapshot}). Once the BH reaches the central region, $v_\text{rel}$ greatly increases up to $100~\mathrm{km~s^{-1}}$, making the accretion rate far below the Eddington value. This suppression of BH growth by SNe is consistent with the previous studies \citep[e.g.][]{Dubois+2012, Latif+2018}.


The larger seed mass model with $10^{6}~M_{\odot}$ shows
the more efficient mass growth.
The Eddington rate is one to two orders of magnitude larger than 
the model with the initial mass of $10^{5}~M_{\odot}$ until $z \sim 14$
(Fig.~\ref{fig_BH_growth}b).
Since a more massive BH can more strongly bind the surrounding gas
against the radiative heating,
the growth rate increases with increasing the BH mass.
After $z \sim 17$, however, the growth rate suddenly decreases 
as the intense star formation activity makes the galactic gas so turbulent
as in the smaller initial mass case.
This result indicates that increase of the initial seed mass helps the mass growth only in the early evolutionary phase.
Even when the BH approaches the galactic center, it hardly grows by accretion due to the large turbulence.


In our simulation, the DCBH does not settle down in the galaxy center at $z=14$: it is still wandering at the outer galactic region, several $100~$pc from the center for all the models studied.
Fig.~\ref{fig_BH_migration}(a) -- (e) shows the time series of the density distribution
around the galaxy center and positions of the BH, noted by the circles bounded by the dashed line.
This model assumes the initial mass of $10^{5}~M_{\odot}$
and the radiative feedback from the BH is taken into account.
The BH approaches the galaxy center at first (panels a--c),
while it passes by and moves away from the galaxy center afterwards
(panels c--d).
Fig.~\ref{fig_BH_migration}(f) shows the trajectory of the BH,
indicating the BH orbits around the galaxy center still at $z=14$.
Since the BH cannot reach the central high density region,
the accretion rate onto the BH stays far below the Eddington rate.

Note that the accretion rate is suppressed not only by the low gas density,
but by a large velocity relative to the galactic gas,
which has been often overlooked in the previous studies.
The large relative velocity arises because
the BHs are created at $\sim1~$kpc away from the galaxy center.
This large separation is one specific feature of the DCBH formation,
that the cloud should avoid the metal enrichment to form a massive seed BH.
The BH is accelerated to the order of the halo's circular velocity $V_\text{c}$, 
which is written as \citep[e.g.][]{Barkana+2001}
\begin{align}
V_\text{c} = 50~\mathrm{km~s^{-1}} \left ( \frac{M_\text{h}}{10^{9}~M_{\odot}} \right )^{1/3} \left ( \frac{1+z}{21} \right )^{1/2},
\end{align}
where $M_\text{h}$ is the halo mass hosting the massive galaxy.
This large relative velocity greatly suppresses the BH growth,
in combination with the turbulent motion of the galactic gas.

Our calculation also suggests that the radiation feedback
from the BH makes its inward migration more inefficient (Fig.~\ref{fig_gas_props_around_BH}c).
This is mainly because the photo-heating disperses the dense gas that accumulates in the downstream of the BH,
which carries away the kinetic energy of the BH otherwise \citep[e.g.][]{SouzaLima+2017}.
Recent high-resolution simulations
reveal that the BH is even accelerated by the gravitational pull from a dense shell created ahead of the BH, 
which further makes the inward migration more difficult \citep{Park+2019,Toyouchi+2020}.

\section{Discussion} 
\label{sec::discussion}
\subsection{Dynamical friction}

Our simulation has shown that the DCBH does not grow in mass during the initial $100~$million years. 
Such inefficient growth mainly arises from the fact that 
the BH cannot migrate toward the galactic center.
The BH continues to wander around the low-density outskirts of the galaxy at a high velocity of several $\times 10~{\rm km~s^{-1}}$. The accretion rate onto the BH, which is basically given by the Bondi rate, is accordingly well below the Eddington value.


We here quantitatively see how the large relative velocity stuns the BH mass growth. The Bondi accretion rate normalized by the Eddington rate $f_\text{Edd}$ is written as
\begin{align}
f_\text{Edd} = 0.032 \left ( \frac{c_\text{eff}}{10~\mathrm{km~s^{-1}}} \right )^{-3} 
\left ( \frac{M_\text{BH}}{10^{5}~M_{\odot}} \right ) 
\left ( \frac{n}{1~\mathrm{cm^{-3}}} \right ),
\end{align}
where $c_\text{eff}$ is the effective sound speed defined as $\sqrt{c_\text{s}^{2} + v_\text{rel}^{2}}$ and $n$ is the surrounding gas density.
Given that $c_\text{eff}$ is equal to the sound speed of the ionized gas $10~\mathrm{km~s^{-1}}$, the mass accretion proceeds at the Eddington rate as long as $n \gtrsim 30~\mathrm{cm^{-3}}$.
However, our results show that $c_\text{eff}$ becomes as high as $100~\mathrm{km~s^{-1}}$ 
both by the turbulence caused by the intense SNe 
and by the acceleration of the BH motion when falling into the galactic potential well (Fig.~\ref{fig_gas_props_around_BH}b).
With this high value of $c_\text{eff}$, the gas density required to attain the Eddington rate is $n \gtrsim 3 \times 10^{4}~\mathrm{cm^{-3}}$, much larger than the typical value in the outer galactic region.
Since such a dense gas is only available in the galaxy center, the BH needs to dissipate the kinetic energy to migrate toward the center for the efficient mass accretion.


One mechanism to bring the BH toward the galaxy center is the so-called ``dynamical friction'', where the BH interacts with the surrounding stars and gas which carry away the BH's kinetic energy. The timescale for the dynamical friction to operate is evaluated as follows. The drag force due to the dynamical friction is analytically written as \citep{Ostriker1999},
\begin{align}
F_\text{DF} = -I \frac{4\pi G^{2} M_\text{BH}^{2} \rho_{\infty}}{v_{\infty}^{2}},
\end{align}
where $\rho_{\infty}$ is the background density,
$v_{\infty}$ is the velocity relative to the background media, and
$I$ is an order of unity non-dimensional parameter,
which depends on the local Mach number $v_{\infty}/c_\text{s}$.
The corresponding timescale can be evaluated as,
\begin{align}
t_\text{DF} &= \frac{v_{\infty}}{F_\text{DF}/M_\text{BH}}, \nonumber \\
&= \frac{v_{\infty}^{3}}{4 \pi I G^{2} M_\text{BH} \rho_{\infty}}, \\
&= 1.7\times 10^3~\mathrm{Myr} \nonumber \\
&\;\;\; \left ( \frac{M_\text{BH}}{10^{5}~M_{\odot}} \right )^{-1} 
\left ( \frac{v_{\infty}}{50~\mathrm{km~s^{-1}}} \right )^{3} \left ( \frac{n}{10^{2}~\mathrm{cm^{-3}}} \right )^{-1}. \label{eq::tmigration}
\end{align}
Since the cosmic age at $z=6$ is $800~$million years,
there is not enough time for the DCBH to sink toward the galaxy center.


\begin{figure}
	\centering
		\includegraphics[width=8.cm]{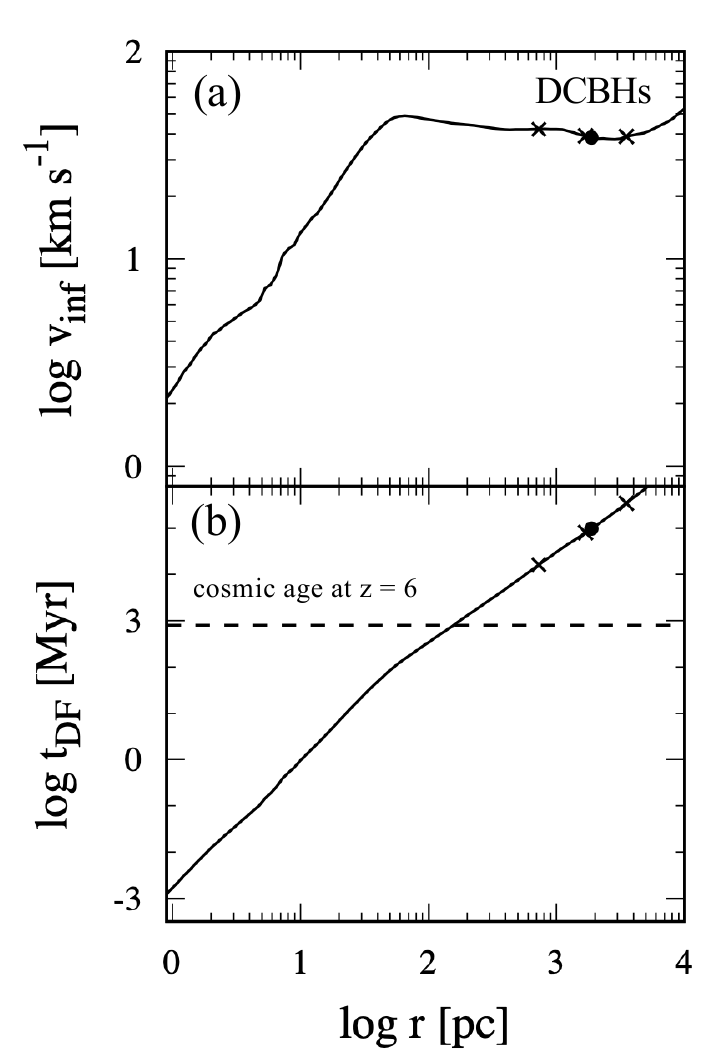}
		\caption{(a) The BH initial velocity $v_\text{inf}$ (eq.~\ref{eq::vinf}) as a function of the distance from the galaxy center. 
		We evaluate $M_\text{enc}$ from the snapshot at $z=20.8$, when the DCBH under consideration is formed.
		The bold circle indicates its initial position in our simulation.
        We also show the positions of three DCBHs that are formed later at $19 < z < 20$ by the cross symbols.
		(b) The dynamical friction timescale $t_\text{DF}$ (eq.~\ref{eq::tmigration}). 
		The dashed line represents the time scale of $800~$Myr, which is the cosmic age at $z\sim6$.}
		\label{fig_tmigration}
\end{figure}

\begin{figure}
	\centering
		\includegraphics[width=8.5cm]{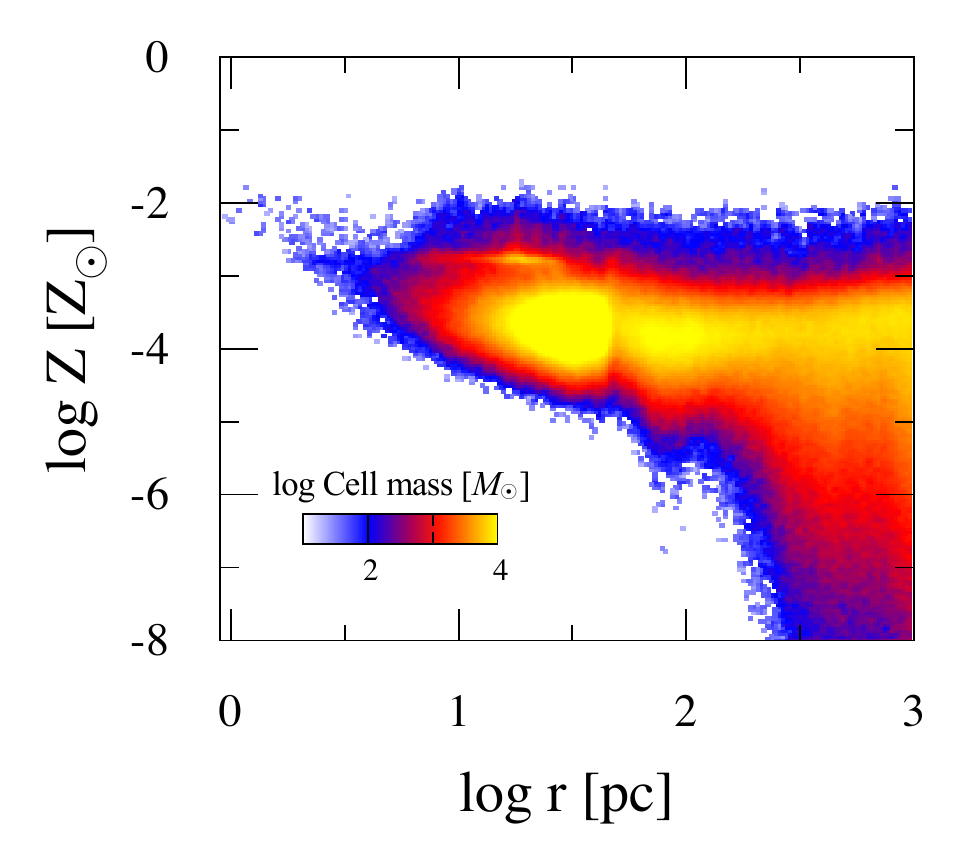}
		\caption{The radial distribution of gas metallicity $Z$ in the DCBH-forming galaxy at $z=20.8$,
		where $r$ is the distance from the galaxy center.
		Colors show the mass in each cell, 
		where we divide the $Z$--$r$ plane by $200\times200$ grid.}
		\label{fig_metals}
\end{figure}

One simple solution to put the BH at the galaxy center 
is creating a DCBH in a shorter distance, near or inside the virial radius of the host halo.
Given that the seed BH is formed at a distance $r$ from the galaxy center, the initial velocity relative to the galaxy center can be estimated as,
\begin{align}
v_\text{inf} = \sqrt{\frac{2GM_\text{enc}(r)}{r}}, \label{eq::vinf}
\end{align}
where $M_\text{enc}(r)$ is the enclosed mass inside $r$.
Fig.~\ref{fig_tmigration}(a) shows $v_\text{inf}$ as a function of $r$ at the birth of the seed BH.
At $r \gtrsim 50~$pc, $v_\text{inf}$ becomes roughly constant at $40$--$50~\mathrm{km~s^{-1}}$.
Assuming $v_{\infty} = v_\text{inf}$ and inserting the density profile into eq.~\eqref{eq::tmigration}, we can evaluate $t_\text{DF}$
as a function of $r$ (panel b). We see that $t_\text{DF}$ monotonically decreases with decreasing $r$, since the density increases toward the galaxy center. The dashed line indicates the cosmic age at $z=6$, showing that the seed BH needs to be created at $r \lesssim 100~$pc in order for the dynamical friction to operate before $z=6$.


Forming a seed BH close to the galaxy center is preferable for the dynamical friction to operate, while such a region is usually polluted with the heavy elements \citep[e.g.][]{Dijkstra+2014, Agarwal+2017}.
Fig.~\ref{fig_metals} presents the radial distribution of the gas metallicity around the galaxy center at $z=20.8$. We see that 
the gas located at $r\lesssim 100~$pc has the metallicity with $10^{-5}$--$10^{-3}~Z_{\odot}$.
The DCBH formation would not occur in such regions, since the ordinary DC model requires the cloud to be purely primordial to form the SMSs.
If the collapsing clouds have such metallicities, the efficient dust cooling induces vigorous fragmentation at the density $n\gtrsim 10^{10}~\mathrm{cm^{-3}}$ \citep[e.g.][]{Omukai+2008, Latif+2016, Tagawa+2020}.
Recent simulation by \citet{Chon+2020} has found that SMSs and thus massive seed BHs can form even when dust-induced fragmentation takes place. Massive gas inflows still efficiently fuel the gas into the central region and massive stars formed there efficiently grow into SMSs.
\cite{Regan+2020} point out that the number density of such metal-enriched SMS-forming sites is comparable to that of the primordial cases, analyzing their cosmological simulation data. 
We here argue that the metal-enriched cases are even more crucial than the primordial ones because they only offer the efficient accretion growth of the seed BHs.

\subsection{Wandering BHs}
Although we have focused on the accretion growth of a specific DCBH, this is not alone in our simulation. After the birth of this DCBH at $z=20.9$, more pristine clouds satisfy the DC criteria; for instance, additional three DCBHs appear during $19 < z < 20$ in the same cosmological volume. In Fig.~\ref{fig_tmigration}, the cross symbols indicate that the formation sites of those three BHs are also $\sim$~kpc away from the galactic center. 
The distance of $\sim$~kpc is characteristic to the DCBH formation because nearer clouds suffer the metal enrichment while the farther ones lack
the required high LW intensity. 
Starting from similar birthplaces $\sim$~kpc away from the galactic center, subsequent accretion growth of the DCBHs is similar among them:
they hardly grow in mass and remain wandering in the outskirts of the galaxy at $z=14$, the endpoint of our simulation.


Since we do not aim to evaluate the total number of the DCBHs ever formed, we have neglected their formation at lower redshifts $z < 19$. 
This means that we underestimate the DCBH population and their radiative feedback against the surrounding gas. However, our conclusion on the mass growth of the DCBHs should not be altered because anyway they hardly grow even without additional feedback. 
Although not examined here, we expect that birth sites of DCBHs formed later are also $\sim$~kpc away from the center of the host galaxies,
similarly to the cases considered above. Consequently, their accretion growth should also be inefficient because they do not readily settle down to the galactic center.


If DCBH formation continues at a similar rate by the epoch of $z=14$, their expected population exceeds that found in the previous studies, where the DC is prevented by the strong tidal force from a massive galaxy \citep[][]{Chon+2016} or by the metal-enrichment caused by the powerful galactic winds \citep[][]{Wise+2019}. This may be a result of enhanced DCBH formation triggered by radiation feedback from pre-existing DCBHs. In fact, one of the four DCBH-forming clouds in our simulation coincides with that found by \citet{Chon+2017}, who put an intense light source by hand at the galactic center and studied the qualitative effects on the subsequent evolution in the same cosmological volume.
\citet{Chon+2017} conclude that the same cloud collapses owing to the aid of compression by the external hot photoionized gas.
We here follow the evolution in a more self-consistent manner, and find that the radiation from the pre-existing DCBH may have
induced a similar evolution.
Moreover, enhanced LW radiation fields owing to the contribution from accreting DCBHs are favored for satisfying the DC criteria and may lead to further DCBH formation \citep{Yue+2014}.


Whereas we have argued that the DCBHs stay in the outskirt of the host galaxies for sometime without moving to the galaxy center,
we expect divergent long-term histories from $z = 14$ to $z=0$.
The timescale of the migration (eq.~\ref{eq::tmigration}) indicates that the BHs settle in the galaxy center within the Hubble time if they enter a high density region with $n\gtrsim 10~\mathrm{cm}^{-3}$.
When multiple BHs happen to gather near the galaxy center, they may form tight binaries, which further shrink via the dynamical friction by the surrounding gas and stellar components.
Mergers of such binaries can be observed by future space gravitational wave detectors such as Laser Interferometer Space Antenna \citep[LISA;][]{Amaro-Seoane+2012} and Deci-hertz Interferometer Gravitational wave Observatory \citep[DECIGO;][]{Kawamura+2011}. 
We follow \citet{Haehnelt1994} to estimate the detection rate of such merger events $\dot{N}_\text{GW}$ as
\begin{align}
\dot{N}_\text{GW} &= \int 4\pi r(z)^{2} N_\text{DCBH} (z) \mathrm{d}r, \nonumber \\
&\sim 1.0~\mathrm{yr^{-1}} \left [ \frac{N_\text{DCBH} (z) }{10^{-3}~\mathrm{Mpc^{-3}}} \right ],
\end{align}
where $r(z)$ is the comoving coordinate distance and $N_\text{DCBH}$ is the number density of the DCBHs. 
Here, we have assumed that all the DCBHs experienced mergers and their signals are detectable. \citet{Chon+2016} suggest $N_\text{DCBH} \sim 10^{-4}$--$10^{-3}~\mathrm{Mpc^{-3}}$ \citep[see also][]{Habouzit+2016,Wise+2019}, resulting in the event rate of $\sim 0.1-1~{\rm yr}^{-1}$. We note that evaluating the DCBH formation rate is challenging and other authors have proposed much lower values, $10^{-7}$ -- $10^{-9}~\mathrm{Mpc^{-3}}$ \citep{Dijkstra+2008,Dijkstra+2014, Valiante+2016}. Detection of the merging signals will put observational constraints on $N_\text{DCBH}$ \citep[e.g.][]{Hartwig+2018}.


If the dynamical friction is inefficient, the BHs may remain wandering in the outskirts of galaxies even at the present day \citep[e.g.][]{Bellovary+2019}. In fact, recent observations suggest that some nearby dwarf galaxies are harboring candidates of the wandering BHs \citep[e.g.][]{Reines+2020, Mezcua+2020}.
Some of them are in the intermediate ($\sim 10^5~M_\odot$) mass range, and our results suggest that they are possible survivors of the DCBHs.
\citet{Inayoshi+2019a} recently study the observational signatures of such BHs by means of numerical simulations.
They show that the BH acquires the gas through a geometrically thick disk,
inside of which the inflow rate becomes significantly lower than 
the Eddington rate owing to convective motions.
The resulting disk luminosity is too small to be observed in the X-ray band \citep[e.g.][]{Zivancev+2020}, while the radio signal can be detectable by ALMA or VLA if the BH orbits around the outer part of the Milky-Way \citep{Guo+2020}.

\subsection{Effects of limited spatial resolution}

Although our current simulations achieve a high spatial resolution below the Bondi radius, it is still unfeasible to resolve very dense gases with $n \gtrsim 10^{3}~\mathrm{cm^{-3}}$ near BHs owing to substantial computational cost. Dealing with such a dense gas is challenging, but that may be vital to consider more efficient accretion growth of a BH, i.e. via the so-called super-Eddington accretion \citep[e.g.][]{Inayoshi+2016,Lupi+2016}. If such rapid accretion takes place, the BH accretes the gas through the so-called slim disk, in which the resulting radiation efficiency is much lower than that we have assumed, $\epsilon=0.1$ \citep[e.g.][]{Watarai+2000, Ohsuga+2005}.
Other effects that potentially help to attain the super-Eddington accretion, which are neglected in our work, are anisotropy of the radiation emitted from the unresolved part of the disk \citep{Sugimura+2017, Takeo+2018}, and metal enrichment of the accreting gas with $Z \sim 10^{-2}~Z_{\odot}$ \citep[][]{Toyouchi+2019}. 
However, we also note that there are also obstacles to achieve the super-Eddington accretion, for instance, the angular momentum of the accreting gas \citep[e.g.][]{Sugimura+2018}. We need further studies considering all these effects consistently with a sufficiently high spatial resolution.


If rapid accretion occurs in reality because of currently unresolved dense gases, that effect may be incorporated
as a boost factor, which recent simulations have often used \citep[e.g.][]{Booth&Schaye2009, Dubois+2015}. However, we expect that dense and cold clouds should concentrate around the galactic center, where the gravity of the galaxy strongly bind the gas. Since our simulations show that the DCBHs do not settle down to the galactic center but keep wandering around the rarefied outskirts, the boosting effect, if any, only plays a minor role for the early accretion growth of the DCBHs.


\section{Summary}  
\label{sec::summary}

We have studied mass accretion history onto a direct-collapse black hole (DCBH) with cosmological simulations, extending our previous studies where we only focused on its formation stage. Since DCBHs are supposed to form in pristine clouds to avoid the metal enrichment, their formation takes place at $\sim$ kpc apart from associating massive galaxies. The BH initially accretes the gas nearly at the Eddington rate, but the radiative feedback suppresses the rate by about two orders of magnitude after a while. Furthermore, the accretion rate stays far below the Eddington rate, even when the BH approaches the galactic center for the following two reasons. One is that the intense supernova activity injects a large amount of energy into the gas, and causes the supersonic turbulence. The other is that the BH accelerates when falling into the galactic potential well, and obtains a large velocity relative to the gas.  Both of these effects significantly reduce the Bondi accretion rate onto the BH, far below the Eddington value. As a result, we only observe the mass growth of $0.6\%$ during the initial $\sim100~$million years after the seed BH formation. The BH is still wandering within the galactic potential at the end of our simulation.


Our analytic estimation shows that it is difficult for the dynamical friction to bring the DCBH to the galactic center before $z \simeq 6$.
This is because the BH has a high initial velocity of a few $\times$ $10~\mathrm{km~s^{-1}}$ when it appears at the birthplace, which is far away from the galactic center to avoid the metal enrichment.
The resulting timescale of the dynamical friction is comparable to or greater than the cosmic age at $z \sim 6$, suggesting that the BH should continue to wander in the low-density outskirts of the galaxy without migrating inward toward the galactic center. 
Therefore, it should be unlikely that such DCBHs eventually grow into the observed SMBHs exceeding $10^9~M_\odot$ before $z \simeq 6$.

One possibility to attain the efficient growth is forming a seed BH much closer to the galaxy center, with the distance of $r \lesssim 100$~pc. This inevitably leads to a slight metal enrichment with expected gas metallicities of $10^{-5} - 10^{-3}~Z_\odot$, which conflicts with a standard condition required by the DC model. However, \citet[][]{Chon+2020} show that the supermassive star formation occurs even in somewhat metal-enriched environments, where the dust-induced fragmentation operates to create numerous stars.  
A massive seed BH formed in such a way migrates to the galactic center promptly, and it may further grow via efficient mass accretion. 
Our results also predict a number of DCBHs that never reach the galactic centers owing to inefficient dynamical friction. They might be still wandering at the outskirts of nearby massive galaxies including Milky-Way, and they are potential targets for future observations.

\section*{Acknowledgements}
The authors wish to express their cordial thanks to Prof. Naoki Yoshida for his continual interest, advice and encouragement.
We also thank Kazu Sugimura for fruitful discussion and comments. This work is financially supported by the Grants-in-Aid for Basic Research by the Ministry of Education, Science and Culture of Japan 
(19J00324: S.C., 17H02869, 17H01102, 17H06360: K.O., 19H01934: T.H.). We conduct numerical simulation on XC50 at the Center for Computational Astrophysics (CfCA) of the National Astronomical Observatory of Japan and on XC40 at Yukawa Institute of Theoretical Physics in Kyoto University. We use the SPH visualization tool SPLASH \citep{SPLASH} in Figs~\ref{fig_snapshot_zoom}, \ref{fig_snapshot}, and \ref{fig_BH_migration}. 

\section*{DATA AVAILABILITY}
The data underlying this article will be shared on reasonable request to the corresponding author.

\bibliography{biblio2}

\bsp	
\label{lastpage}
\end{document}